# Considerations regarding security issues impact on systems availability

Emil Pricop, Sanda Florentina Mihalache, Nicolae Paraschiv
Automatic Control, Computers & Electronics Department
Petroleum-Gas University of Ploiesti
Ploiesti, Romania
e-mail: emil.pricop@upg-ploiesti.ro; sfrancu@upg-ploiesti.ro; nparaschiv@upg-ploiesti.ro

Jaouhar Fattahi
Dept. of Computer Science & Software Engineering
Université Laval
Québec, Canada
e-mail: jaouhar.fattahi.1@ulaval.ca

Florin Zamfir
Automatic Control, Computers & Electronics Dept.
Petroleum-Gas University of Ploiesti
Ploiesti, Romania
e-mail: florin.zamfir@upg-ploiesti.ro

*Abstract* – Control systems behavior can be analyzed taking into account a large number of parameters: performances, reliability, availability, security. Each control system presents various security vulnerabilities that affect in lower or higher measure its functioning. In this paper the authors present a method to assess the impact of security issues on the systems availability. A fuzzy model for estimating the availability of the system based on the security level and achieved availability coefficient (depending on MTBF and MTR) is developed and described. The results of the fuzzy inference system (FIS) are presented in the last section of the paper.

*Keywords- availability, security, availability coefficient security level, security impact*

## I. INTRODUCTION

Control systems are key components of various critical infrastructures such as energy production and distribution facilities, petrochemical plants, pharmaceutical plants or weapons factories [1]. Without any doubt, the control systems represent elements that are responsible for the good functioning of the critical infrastructures.

There are many facets that should be taken into account when analyzing the control systems existing in critical infrastructures. The most significant facets, defined along with others in [2], are: technical performance, reliability, availability, maintainability, safety and security. These systems attributes are not independent, each of them affecting the others.

The technical performances can be evaluated according to the concepts from systems theory and indicated by the overshoot and the transient time values.

Reliability is defined as the system's ability to perform its functions on a given time interval. In the current practice reliability is regarded as an integrating concept that includes both availability and maintainability.

Availability, the main parameter analyzed in this paper, is defined in strict relation to reliability. Availability can be defined as the ability of the system to perform its functions on at a given time.

Maintainability is a parameter that intervene in the estimation of achieved availability [3], showing the amount of effort required to restore an equipment or system to working state in case of a fault.

Safety, as stated in [4] consist in the ability of the system to not harm people when any kind of error or fault occur in system functioning.

On the other side security represent the system ability to not produce any type of loss or destruction when a deliberate malicious act occurs. The two concepts *safety* and *security* should be clearly separated and analyzed according to their differences.

Safety and security are critical aspects impacting the system functioning. A system having a good safety and reliability levels, but lacking security will produce surely economic losses for its owner in care of a cyber-attack or related incident. On the other side a security hardened system with very little reliability will not be useable. The main challenge for academia and industry is to find the equilibrium between functionality, security and reliability, in other words to assure a good security level without affecting the system efficiency.

It is obvious that security incidents will affect negatively the system availability by interrupting its correct functioning. The basic security concepts associated with control systems will be presented in the second section of this paper. In the third section of the paper we will present the state of the art in estimating the system availability. The last part of the paper is dedicated to designing and implementing a fuzzy inference system for assessing the security incidents impact on system availability. The authors discuss the obtained results and define a new method to define the availability coefficient, starting from achieved availability, based on MTBF – mean time between failures – and MTR – mean time of reparation, and taking into account the systems' security level.

## II. CONTROL SYSTEMS SECURITY BASICS

Control systems are nowadays key components of industrial infrastructures. Their main role is to implement algorithms and methods required to monitor and keep at a desired level one or more process parameters without human intervention.

The simplest control system comprises at least one controller, having processing power, one sensor and one actuator. All those elements transfer information between them and with the external environment.

Starting from the simplest control systems, there were developed more complex systems integrating multiple controllers in a hierarchized approach. It is obvious communication between controllers is critical in this situation.

The recent developments in control systems allow connecting them to the economic and production planning facilities such as ERP or CRM applications. The economic objectives generated at a higher level are transposed in technical objectives that can be handled by control systems. Also these systems can be remotely operated and are connected in various industrial networks or even to the Internet, becoming targets for attackers.

Taking into account these considerations it is obvious that control systems are threatened by various attackers that can exploit their vulnerabilities. A secure system lacks the vulnerability to sabotage, tampering, viruses, worms or informatics attacks [1], [5].

In order to evaluate the security level of a system a threat – vulnerability analysis should be completed. It allows identifying the threats – attackers – and vulnerabilities – security breaches that can be exploited by the attacker in order to gain access to the system or to interrupt its functioning.

The main vulnerabilities associated to control systems are presented in [1]. Some attacker profiles are described in [5].

The main security concern related to control systems is not data confidentiality, but it is data and system availability. This approach is different from classical information security where data confidentiality and integrity were critical issues and it is justified by the impact of an interruption in control system functioning. For example, a control system that regulate the temperature in a distillation column from a refinery is critical for plant functioning; its failure will lead in a costly controlled installation shutdown procedure or in a worse scenario to an accident with high impact on people and environment.

The control systems availability is clearly affected by the security incidents and it should be related to the systems' security level. In the following section, a state of the art in systems availability estimation taking into account security is presented.

## III. STATE OF THE ART

In reference [6] is defined the availability coefficient, named in this paper achieved availability, $K_D$ by the following relation:

$$K_D = \frac{MTBF}{MTBF + MTR} \cdot 100 \quad (1)$$

where MTBF represent Mean Time Between Failures and MTR is the Mean Time Reparation. The security issues / attacks that produce interruption in system functioning are not quantified by (1). This relation contain only terms related to faults that occur normally during the operation period of any equipment.

It is obvious that relation (1) does not show explicitly the impact of security issues on systems availability. A cyber-attack may interrupt the system functioning, interruption that is not quantified nor by MTBF or MTR. Much more, other attack types such as information hi-jacking will not interrupt the system functioning by shutting it down, but will not permit the system to attain the desired state, leading to incorrect functioning which is equivalent to a decrease in the real availability of the system.

In following paragraphs, the authors present their own literature study focused on estimating systems availability when security incidents happen. From our study we can conclude that the researches focus on wireless sensor networks – WSN node availably.

Oreku [7] proposes in his work a mathematical approach based on probability for calculating the reliability/availability coefficients associated to WSNs (with different architectures). The study concludes that reliability for security is dependent on three elements: connectivity, network parameters and Data aggregation.

Bandirmali and Erturk [8] create a WSN link layer security protocol and emphasize the role of reliability, usability and performance as constraints in its design. Reliability is a critical parameter for wireless sensor network based applications where data integrity and confidentiality should be assured. The study does not provide a mathematical approach for calculating the reliability coefficients.

An interesting technique for evaluating the reliability of WSN with star, cluster and mesh type topologies is proposed in Distefano [9]. It is proposed a method to assess the system reliability through the dynamic reliability block diagrams modelling approach. The modelling approach has two parts: the dynamic approach of each node, taking into account the power consumption in standby, interference existence and the static part that focus on WSN topology.

Mahmoud et al. [10] analyze reliability in various topologies and protocols for wireless sensor networks. The main focus of their research is proposing a 3D reference model. The authors do not focus specifically on security impact on system availability, but they analyze the retransmission issues in wireless communication, in connection-oriented and link-oriented environments.

The results found in Jang et al. [11] provide a knowledge base in the area of sensor hardware design, field data collection, WSN application strategy and network topology design. Jang et al conclude that the different characteristics of construction materials may affect the level of reliability performance in wireless communication.

Stavrou and Pitsillides [12] focus on presenting the security impact on critical infrastructures correct functioning. In their research they analyze the security of various routing protocols for wireless sensor networks and concludes that the next protocols that will be developed and used should be designed having security as the main criteria.

Ayday and Fekri present in [13] an innovative node to network broadcast algorithm that ensure very good communication system security. The proposed solution assure system functioning even when an attack is deployed and some nodes in the wireless network are compromised, leading to a good system reliability and availability.

Parthasarathy et al. [14] present the challenges concerning reliability, flexibility and security supported by a WSN deployed in hazardous environments (remotely controlled WSN in a volcano area).

In reference [17] Cazorla et al. focus on critical infrastructures security. First the authors show that critical infrastructures are key components of all the system involved in current human activity. Cazorla et al. highlights the threats and vulnerabilities of such systems and then, based on their analysis, indicate requirements imposed for creating a protection layer for critical infrastructures. They developed a framework that allow evaluation of an Intrusion Detection System for critical infrastructures based on various criteria such as impact on system availability, performance and specific security metrics.

Some researchers developed fuzzy or genetic algorithms based methods for reliability evaluation of repairable systems. Such approaches are presented in [18], [19], [20], [21] and [22]. The fuzzy approach is used in order to model the failure rate (MTBF) and mean reparation times (MTR) instead of using classical methods, in order to overcome the complexity of the probabilistic approach.

## IV. FUZZY SYSTEM PROPOSAL AND RESULTS

Estimating the system availability while taking into account the security incidents impact is a very challenging task. Since there is some vagueness level in the description of the system availability a fuzzy inference system is appropriate in order to describe the system behavior using a linguistic method.

In this section of the paper, a fuzzy approach on modelling the security impact on system availability is presented.

There are two parameters that should be taken into account: $K_D$ – achieved availability coefficient (1) and $K_S$ – security level coefficient.

The security level coefficient shows how secure is a given system. If $K_S$ has a value close to 1, it indicates that the system is very secure and it is hard to be affected by attackers. If $K_S$ has a very low value, close to 0, it shows that the system is prone to attackers.

TABLE I. PROPOSED RULES FOR AVAILABILITY ESTIMATION

| No. | $K_D$ | $K_S$ | Global availability |
|---|---|---|---|
| 1 | Very Small | Very Small | Very Small |
| 2 | Very Small | Small | Very Small |
| 3 | Very Small | Medium | Very Small |
| 4 | Very Small | Big | Very Small |
| 5 | Very Small | Very Big | Very Small |
| 6 | Small | Very Small | Very Small |
| 7 | Small | Small | Small |
| 8 | Small | Medium | Small |
| 9 | Small | Big | Small |
| 10 | Small | Very Big | Small |
| 11 | Medium | Very Small | Very Small |
| 12 | Medium | Small | Very Small |
| 13 | Medium | Medium | Medium |
| 14 | Medium | Big | Medium |
| 15 | Medium | Very Big | Medium |
| 16 | Big | Very Small | Very Small |
| 17 | Big | Small | Small |
| 18 | Big | Medium | Medium |
| 19 | Big | Big | Big |
| 20 | Big | Very Big | Big |
| 21 | Very Big | Very Small | Very Small |
| 22 | Very Big | Small | Small |
| 23 | Very Big | Medium | Medium |
| 24 | Very Big | Big | Big |
| 25 | Very Big | Very Big | Very Big |

In order to implement the fuzzy inference system for global availability evaluation the two inputs $K_D$ and $K_S$ are described by 5 values: "Very small", "Small", "Medium", "Big" and "Very big". Given this linguistic description there are 25 rules that can be defined for the proposed fuzzy system. The rules, which are presented in Table 1, are constructed based on authors experience with security assessment of the systems.

The output of the proposed fuzzy inference system is the global availability coefficient which shows the cumulative effect of common failures (indicated by $K_D$ value) and security issues (indicated by $K_S$). Rule number 25 shows that a system with good security ($K_S$ – very big) and good reliability will have a very good global availability coefficient.

The fuzzy inference system generates a 3D surface shown in Figure 1. The global availability is defined as f($K_D$, $K_S$) and its value can easily be determined for each $K_D$ and $K_S$ values.

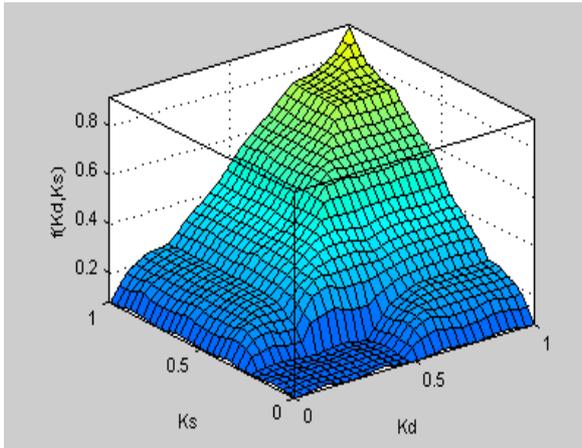

Figure 1.   Global system availability

The generated surface shows the evolution of the global availability when $K_D$ and/or $K_S$ modify.

The cases of two systems with different security levels will be analyzed in the following paragraphs.

In order to easily understand the FIS results in Figure 2 is presented the same surface as in Figure 1 using level curves.

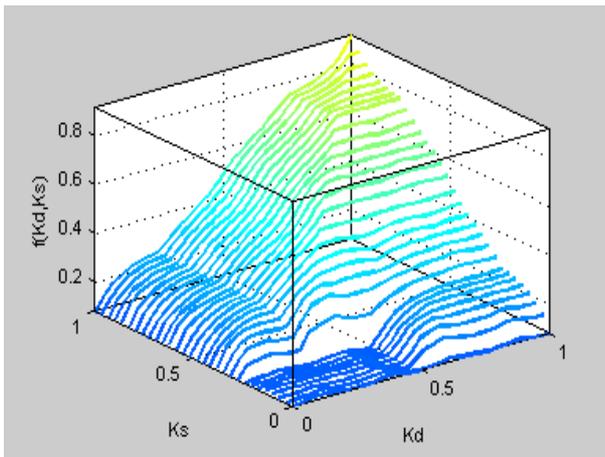

Figure 2.   Global system availability (level curves)

There is an ideal case when $K_D$=1 and $K_S$=1 lead to a global availability coefficient $K_G$ =1, which is impossible in a real situation, since such a system is not realizable. A value of $K_D$=1, means that MTR (reparation time) is zero and this is an impossible situation. The case of $K_S$=1 correspond to an invulnerable system from the security point of view which is also impossible.

Based on the results shown in Figure 2 there can be built some scenarios for results analysis. The first case is of a system with good security. It will have a security level $K_S$=0.75. The global availability will depend on the achieved availability as shown in Figure 3.

The global availability value will not pass 0.75, since the security incident that may occur will decrease the achieved availability ($K_D$). The dependence on system reliability is in this case almost linear.

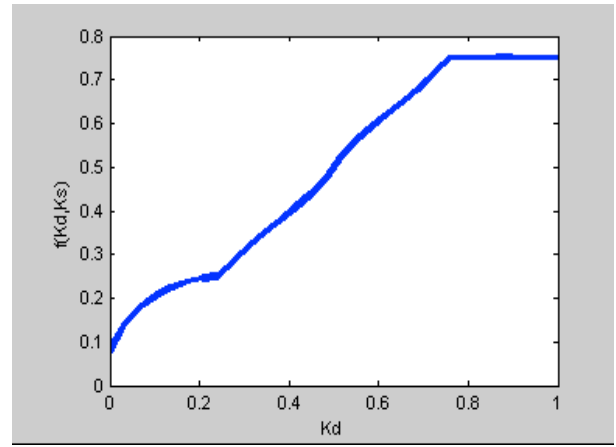

Figure 3.   Global system availability for a system with $K_S$=0.75

In figure 4 is presented the case of a system with a very vulnerable system. Its security level $K_S$ is only 0.25, showing that there are taken only basic security measures such as a general firewall or a standard antivirus protection.

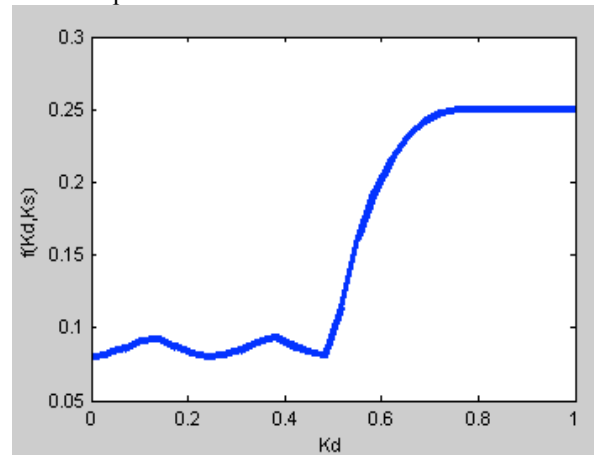

Figure 4.   Global system availability for a system with $K_S$=0.25

Global availability coefficient varies in the case presented in Figure 4 between a minimum value of 0.08 when $K_D$ is close to 0 to a maximum of 0.25 when $K_D$ is over 0.8. The saturation present in the right side of the graph shows that the global availability coefficient is limited by the security level, so a system with a good achieved availability, but with a large number of security issues will not have a satisfying global availability.

From the previous analysis of graphs shown in figures 1-4 it is obvious that equilibrium between achieved availability and security level should be found in order to maintain the protected system in a good state.

## V.   CONCLUSION

It is obvious that security incidents should be quantified when analyzing the control systems availability. In the context of rising threats on control system the achieved availability, based on MTBF and MTR, is not enough for evaluating the actual system availability.

In this paper the authors propose a fuzzy inference system that estimates the global availability taking into account two parameters: achieved availability and security level.

The proposed fuzzy inference system permits to analyze the relation between achieved availability, defined in a conventional way, and security level of a given system. It is only one of the first steps necessary in order to build a framework for evaluation of control systems security.

The main challenge is to build a framework that will allow both academia and industry to find the equilibrium between functionality, security and reliability, in other words to assure a good security level without affecting the system efficiency.